# Corporate Governance and Firms Financial Performance in the United Kingdom

**Martin Kyere [(1,2)] and Marcel Ausloos [(2,3,*)]**

[1] **Aston University, Aston Business School,
Birmingham, B4 7ET, United Kingdom**

[2] **University of Leicester, School of Business,**

**Leicester, LE2 1RQ, United Kingdom**

[3] **Bucharest University of Economic Studies,**

**Department of Statistics and Econometrics,
Bucharest, Romania**

[(*)] **corresponding author: marcel.ausloos@uliege.be**

**ORCID: 0000-0001-9973-0019**

We confirm that this article contains a Data Availability Statement (see section 3.1).

We confirm that we have included a citation for availability of data in the references section (see LSE, 2014).

**Abstract**

The objective of this study is to examine empirically the impact of good corporate governance on financial performance of United Kingdom non-financial listed firms. Agency theory and stewardship theory serve as the bases of a conceptual model. Five corporate governance mechanisms are examined on two financial performance indicators, return on assets (ROA) and Tobin's Q, employing cross-sectional regression methodology.

The conclusion drawn from empirical test so performed on 252 firms listed on London Stock Exchange for the year 2014 indicates a positive or a negative relationship, but also sometimes no effect, of corporate governance mechanisms impact on financial performance. The implications are discussed.

Thereby, so distinguishing effects due to causes, we present a proof that, when the right corporate governance mechanisms are chosen, the finances of a firm can be improved. The results of this research should have some implication on academia and policy makers thoughts.

**1. Introduction**

The aim of this study is to examine the impact of "good" corporate governance on financial performance of firms in the United Kingdom. Turnbull (1997) defines corporate governance as all the influences affecting the institutional process, including those pointing to the controllers and/or regulators, involved in organising the production, sale of goods and services. According to Ehikioya (2009), corporate governance is concerned with processes and structures through which members interested in the firm take active measure to protect stakeholders' interest.

Corporate governance has become more relevant in contemporary times as companies grow and expand both in developed and emerging economies (Freeman, 1983, 2010). As companies expand, they use local raw materials, employ local workforce, sell to the community, pay taxes, etc., that supposedly benefit the community. In addition, recent corporation scandals have been blamed mainly on "bad" corporate governance. (It is almost a daily occurrence to hear news upon scandals ruining corporations.) Consequences of firms' failure are huge; they can be felt in every aspect of society. For instance investors' capital can be wiped out overnight, job losses can occur, etc. (Mallin, 2016).

There is another side to the story: interest groups known as stakeholders' activities can also affect the corporation. For instance, if some society is discontent with the operations of the corporation, it may react negatively towards the firm. Thus, one can boycott its products. As a result, companies may modify their "usual governance", now focusing on social friendly issues departing from idea of shareholders primacy, - when activities are mainly geared toward maximising shareholders aims (Rodriguez-Fernandez, 2016). In addition, there is some evidence to suggest that investors are willing to pay high premium for shares of firms perceived to have a good corporate governance structure (Clarke, 2007). This affirms why corporate governance mechanisms can be considered related to the financial performance of firms.

Over the past decades, there have been many academic researchers investigating links between corporate governance and firm financial performance. Most of these academic researches point out that good corporate governance has a positive impact on firm's financial performance (Stanwick and Stanwick, 2002); however, other researchers have a different view (Donaldson and Davis, 1991; Jensen and Meckling, 1976; Ehikioya, 2009).

Firms require investors' funding to undertake expansion projects. There is evidence to suggest that corporations that improve good corporate governance mechanisms are able to increase the firm's value by 10% to 12% (Stanwick and Stanwick, 2002). The argument is that, before investors think of investing in corporation, they take into consideration the firm corporate governance mechanisms. According to Weir (1997), a firm for which corporate governance structure is seen as "undesirable" has to struggle to get loans, for example. Mallin (2016) points out that before investors commit their funds to investment activity, they consider indicators like insider shareholder, audit committees, board independence, board size, CEO duality, etc., all related to the corporate structure of the firm. In response, firms are now begun to design programmes of good corporate governance that would be attractive to providers of funds.

Yet, according to Cadbury (2000), corporate governance difficulty arises because of separation between shareholders of the business and its control in response to a system by which corporations are directed and controlled. Sometimes, an agent (manager) may have some opposing interest to that of the principal (shareholder) (Jensen & Meckling, 1976). The problem of conflict of interest can occur as a result of asymmetric information resulting from imperfect contractual agreement between managers and shareholders. Such an information can serve as an incentive to managers to pursue self beneficial business projects at the detriment of shareholders. In addition, the board of directors may find that their business interests collide with their fiduciary duties.

One role of corporate governance is to manage these conflicts between the principals and the agents. Good corporate governance, therefore, should have strong internal mechanisms to manage various interest groups, whence to reduce high agency cost; this was of course discussed already along time ago by Rose-Ackerman (1973) or by Fama (1980).

In the post–Enron financial turmoil in Asia and WorldCom in USA, there was a shift of corporate governance focus from its traditional grounds of agency conflicts to ethical issues such as accountability, transparency, disclosure, and reporting. The public demand for corporate accountability, following the high profile corporate scandals stimulated policy makers, academics, and public/private sectors to strengthen the effort of good corporate governance in corporations (Mallin, 2016). According to Aguilera (2005), the Sarbanes-Oxley Act 2002 for instance was enacted in USA in light of institutional contingencies to ensure that boards of directors adhere to best practices of corporate governance guidelines such as disclosure and honest reporting in corporations. Good corporate governance is to be centred on core principles of accountability, transparency, fairness and responsible management.

Addressing these concerns through business decision-making process has not only benefited investors, but also employees, consumers, and communities by strengthening their voices at general assembly meetings (Gill, 2008).

However, the recent financial crises in 2008 has reinvigorated the debate again as to whether good corporate governance (positively or negatively) influences firms financial performance at all. To help provide unbiased judgement into good corporate governance and the impact on firms' performance, we will research corporate governance mechanisms in UK firms. A random sample of UK firms listed on London stock exchange for 2014 will be selected, thereby avoiding sectorial bias.

A complete theoretical framework based on agency and stewardship theories will aid in answering the research question (Donaldson and Davis, 1991). Corporate governance mechanisms such as insider shareholder, board size, board independence, CEO duality, and Audit committee meetings will be used in the study. This is in line with studies by Ehikioya (2009) and Christensen *et al.* (2010). Financial and market performance of the firm will be here captured using following proxies, respectively: (i) Return on assets (ROA); (ii) Tobin's Q (Perfect and Wiles, 1994; Tejersen et al., 2016). These variables will be controlled using firm's size and leverage. We will test a random sampling of 252 firms listed on London Stock Exchange, different from other prior studies in which the sample is mostly picked-up from FTSE100 companies.

Our report is structured as follows: here below, Section 2 contains some literature review. Section 3 outlines the methodology indicating hypotheses and a description of the variables. Section 4 contains the description of the quantitative results so obtained with a statistical significance discussion. Section 5 translates such findings into practical considerations, examining all variables. The final section (6) is reserved for concluding remarks and recommendation about future research directions.

## 2. Literature Review

### 2.1. Theoretical Background

It has been recalled that a difficulty for nowadays implementing some "good" corporate governance resides in the possibly conflictual relationship between the shareholders and the board of directors. This has been addressed by both agency theory and stewardship theory (Donaldson and Davis, 1991).

### 2.1.1. Agency Theory

The agency theory details the relationship between the managers (agents) and the shareholders (principals) (Donaldson and Davis, 1991). It seeks to resolve divergent interests between management of the organisation and the owners, prescribing ways of resolving such conflicts, like delegating a decision-making authority to the agents who manage a project.

Along the agency theory, corporations stand a chance to increase financial performance if cost is minimized. The agency cost can be seen as a value loss by shareholders because of divergence in interests of managers and owners (Jensen & Meckling, 1976). In addition, agency costs are captured in the stock market that affects the company's share prices. Therefore if agency cost is properly managed, it can help for improving shares value, that is, it improves the overall financial performance of the firm. According to Jensen and Meckling (1976), agency costs is measured as the sum of monitoring costs, bonding costs and residual costs. Therefore, in order to reduce the agency cost the corporate governance mechanism should unravel causes of these conflicts, whence the need for grasping the "agency theory". The effective corporate governance mechanisms control should encourage managers to act in the best interest of the principal (Allen and Gale, 2001).

There is an assumption in the agency theory that, where there is a well-developed market, corporate controls are absent. The consequences lead to market failures, non-existence of the markets, moral hazards, asymmetric information, incomplete contract and moral selection. Various studies however, have suggested that proper monitoring, healthy market competitions, control of executive pay, prudent debt sourcing, efficient board of directors, markets for corporate control and concentrated holdings can help resolving the agency problem (Bonazzi and Islam, 2007). The supporters of agency theory argue that, the role of CEO and chairperson should be assigned to separate individuals. This will ensure proper check and balances between CEO and the chair person (Gillan, 2006).

**2.1.2. Stewardship Theory**

Unlike the agency theory that suggests that the role of CEO and chairperson should be separated, the stewardship theory argues that both roles should be combined. The stewardship theory suggests that directors are able to achieve organisational objective of shareholders by maximising their utility rather than self-serving. Some available empirical evidence supports the side of this argument of stewardship theory (Donaldson and Davis, 1991).

Moreover, stewardship theory predicts that allowing managers to work with discretion can encourage them to work better. Scholars on this side of the debate concur that managerial behaviour is not only driven by financial reward but also requires discretion to enable them to maximise the shareholders' value. In addition, stewardship theory stresses that the concern of managers for their reputation and their career intended progression compel them to act in the interest of shareholders; therefore, agency cost will be minimized (Donaldson and Davis, 1991). There is a psychological side of the argument that managers are able to give up their best when they have job satisfy. Clarke (2004) points that allowing managers to take decisions on their own without having to go through bureaucratic processes improve job satisfaction that contributes towards the overall financial performance of the firm.

Besides, Fama and Jensen (1983) agued that managers have greater access to specific insider information, about the going concern of the organisation, than independent directors. Therefore, managers are expected to have acute knowledge of the operations of the company that will help them make well informed decisions. In that line of thought, the stewardship theory suggests that a low number of independent directors is ideal for companies (Donaldson and Davis, 1991; Christensen et al., 2010). In addition, the stweardship theory affirms that insider-dominated board of directors is more effective in achieving the organisational objective because of finer accessibility to information and technology. Finally, the stewardship theory maintains that the CEO essentially wants to work well rather than opportunistically exploits the system, - as also suggested by the agency theory (Donaldson, 1990).

## 2.2. Empirical Framework

### 2.2.1. Insider Shareholder

Insider shareholder is a term used to describe a director or senior officer of a corporation who owns some shares of a corporation, - usually more than 10% of the voting shares (Jensen and Meckling, 1976). According to Jensen and Meckling (1976), the size of the shareholding by the insider has effects on the general financial performance of the firm. Jensen and Meckling (1976) observed that a rise in insider shareholding by insiders does reduce the agency cost. The logic behind this finding was that managers who own significant shares of the company would not invest in destructive or excessive high-risk projects. Therefore, by principle, managers will prudently invest in projects that are likely to reap high returns.

Several studies show indeed that increasing the proportion of insider shareholding beyond an optimal point reduces financial performance. For instance, Fama and Jensen (1983) point that such an increase can result in managerial entrenchment. Recently, Gupta and Sachdeva (2017) tested a comprehensive data set of hedge funds on financial performance of firms with much or little insider shareholding, - using multiple linear regression models. It is found that firms with insider investment perform better than others. The findings also support the view that increases in insider shareholding, up to an optimal point, about 20% of shares, (could) increase returns.

McConnell and Servaes (1990) had also found that an increase in insider shareholder increases the firm's performance, but beyond 40% to 50% a decline in firm's performance occurs. Yet, Agrawal and Knoeber (1996) reported that, insider shareholder predictive effects disappear when additional corporate governance mechanisms are included - in a single ordinary least square (OLS) regression.

2.2.2. Board Size

The theories of economics show that the board of directors plays an important role in the corporate governance structure of corporations (Fama and Jensen, 1983). The concern of shareholders has to do with whether the board of director is capable to monitor/control managers to act in the interest of the owners. The general notion is that companies that have a large board size are likely to have effective supervision that can improve firm performance. Anderson et al. (2004) and Williams et al. (2005) argued that a large board is likely to possess specialised skills prerequisite for efficient toward better performance. Haniffa and Hudaib (2006) also obtained a positive relationship between board size and financial performance.

Another hypothesis about a small board size inducing a better performance has been presented by researchers arguing that limiting a board size rather improves communication and decision-making (Lipton and Lorsch, 1992; Jensen, 1993; Yermack, 1996; Christensen et al., 2010; Akshita and Sharma, 2015). Lipton and Lorsch (1992) suggested that a board member number should not exceed 10. Yermack (1996) discovered an inverse association between board size and market valuation measured by Tobin's Q. In this respect, Akshita and Sharma (2015) discovered an interesting finding that a large number of board directors is considered to be an expensive affair for a firm, thus affecting firm's performance.

2.2.3. Board Independence

Both agency theory and stewardship theory predict different outcomes depending on the board composition.

According to the agency theory, the board of directors can monitor effectively if these are independent from the management (Fama and Jensen, 1983; Beasley, 1996; Christensen et al., 2010). The argument is that incentives exist for outside directors to protect their reputation that motivate them to exercise decisional control (Fama and Jensen, 1983; Christensen et al., 2010). Beasley (1996) argues that where there are non-executive directors on the board financial statement fraud is unlikely to occur.

Yekini et al. (2015), employing content analysis and panel data set from UK FTSE350 companies, discovered a significant relationship between board independent and information disclosure measured by the proportion of non-executive directors. Their research shows that firms with non-executive directors are more likely than others to disclose information which can improve company performance. Rosenstein and Wyatt (1990) argued that the proportion of independent directors has a positive impact on company's share price and financial performance. Both Yekini et al. (2015) and Rosenstein and Wyatt (1990) support the view of agency theory that non-executive directors can improve company performance because of ability to monitor managers.

In contrast, the stewardship theory argues that inside directors have in-depth knowledge of the company which makes them aware of valuable resources that improve firm performance (Donaldson, 1990).

Other scholars argue in support of stewardship theory that, inside directors are trustworthy stewards of firms' resources and improve company performance because of information asymmetry (Donaldson and Davis, 1991; Nicholson and Kiel, 2007). Agrawal and Knoeber (1996) and Klein (1998) discovered a significant negative association between the number of independent directors and performance of firms.

2.2.4. CEO duality

In some companies, a CEO may have two functions; he serves as chairperson of the board of directors and as executive manager (Elsayed, 2007). Corporate governance guidelines presume that when a CEO is also the chairperson of the board, this leads to concentration of power (ASX, 2007). The primary concern of CEO duality is that, managerial domination of the board of directors can lead to dubious control of meeting's agenda (Firstenberg and Malkiel, 1994). In this regard, CEO/chair may decide to send information that serves personal interest only to the board of directors. Consequently, in corporations where there is a lack of strong monitoring of corporate governance mechanism, management can rather pursue their self-interest (Fama and Jensen, 1983).

According to Lorsch and MacIver (1989), the duality of CEO is a hindrance to board independence, thereby making oversight governance mechanism ineffective. Other studies have discovered some improved company performance when the roles of CEO and the chairperson are separated. Rechner and Dalton (1991) documented that firms opting for independent leadership consistently outperformed those relying upon dual CEO, after testing 141 US firms between 1978 and 1983 adopting longitudinal analysis. Balatbat et al. (2004) examining 313 Australian firms between 1976 to 1983 using multiple linear regression analysis discovered similar result: firms with duality of CEO perform worse than others having no such a duality.

In contrast, the supporters of stewardship theory maintain that duality of CEO/chairperson should rather lead to superior firm performance. Stoeberl and Sherony (1985) argue that duality of CEO allows clear-cut leadership direction for strategy formulation and implementation, that is good for business. In addition, other scholars have said that because powers reside in one person uncertainty with regards to the identity of the person taking responsibility of decision is reduced (Christensen, et al., 2010). Therefore, companies can achieve superior performance when there is duality of CEO. Cannella and Lubatkin (1993) documented a positive association between CEO duality and ROE. Boyd (1995) and Essen et al. (2013) came out with same conclusion.

2.2.5. Audit committees

The role of the audit committee is to ensure that the integrity financial reporting of the corporation meets corporate governance council standard. It also ensures compliance of entities such as mandatory disclosures (Davidson et al., 2005)

Kent and Stewart (2008) discovered that the quantity of disclosure was positively related to frequency of board and audit committee meetings held.

However, there is some conflicting evidence from other scholars work. Klein (1998) discovered that the presence of audit committees do not have any effect on the quality of accounting performance measures. Vafeas and Theodorou (1998) also find no evidence to support that a relationship exists between performance and the "board structure (director affiliation and ownership, chairman affiliation, and committee composition)".

## 2.3. Overview of previous studies relevant to UK

Concerning corporate governance mechanisms on UK's corporation, there are mixed results. Guest (2009) indicated that board size has a strong negative impact on profitability, Tobin's Q, and share returns. According to Guest (2009), UK boards play a weak monitoring role; therefore any influence of large board size is likely to reflect the malfunction of the advisory board. In short, Guest's study supports the argument that a large board size is a hindrance to good communication and effective decision making.

Florackis (2005) discovers the existence of "non-linear impact of managerial ownership and managerial compensation on company performance". He finds a strong evidence that managerial ownership and managerial compensation can work as alternative mechanisms in mitigating agency costs and, therefore, generating good financial performance.

Weir et al. (2002) analyzed "the relationship between internal and external corporate governance mechanism on performance of UK firms within the context of Cadbury Committee's Code of Best Practice". They discovered a "weak relationship" and documented that there is no evidence to support that "firms on top or bottom performance deciles have different corporate governance characteristics". Weir et al. (2002) also raised an argument that it will not be right to impose a corporate governance mechanism on a firm given that market for corporate control is known to be a set of effective means for reducing the agency cost. The Weir et al. (2002) work supports the view that CEO shareholding can cause entrenchment resulting in poor firm performance.

Mura (2007) also documented a weak relationship between non-executive director shareholding and firm performance. But, he discovered that, the size of independent directors have a positive impact on firm performance.

Recently, Al-Najjar (2017) discovered that board size and board independence have a significant impact on the pay of CEO and firm performance for several UK firms.

## 2.4. "Conclusion"

This section 2 covers the pertinent literature of scholars who have done research about corporate governance. However, after reviewing empirical studies in various countries across different years and with different methods, we have to admit that one finds mixed conclusions; there are many disagreements among scholars. Given the expected relevance of good corporate governance, it seems necessary to conduct a further study, if not to clearly unravel the controversies, at least to establish what relationship exists between good corporate governance and the financial performance, for not fully studied specific firms, especially after the recent financial crisis. In order to do so, we consider a sample of firms listed on the London Stock Exchange, examining insider shareholding, board size, board independence, duality of CEO, and audit committees effect on financial performance, variables which appear to be the most crucial ones.

## 3. Methodology

In this section, we illustrate the statistical analysis of empirical data for the variables and their indicator involved in this study. We use multiple regression models with research hypotheses.

## 3.1. Timeframe and Statistical Analysis Model

A sample of firms listed on London Stock Exchange (LSE) for the year 2014 is analysed here below (LSE, 2014)[1]. We have chosen 2014 because the year contains much financial information prerequisite for a robust study. Unfortunately many pertinent data were not documented by companies for the years 2015 and 2016. The periods prior to 2014 have seen a recovering stage for businesses after the recent financial crisis in 2007-2008. This is consistent with prior literature selection of firms, but further interesting due to unique financial characteristics for those years. In accordance with the relevant literature, discussed in Section 2, a multiple regression analysis is employed to help capture the multiple variables involve in the study.

In addition, we use cross-sectional regression analysis to test empirical data, again in accordance with prior literature as this study is for one year; see Rodriguez-Fernandez (2016) and Watsham and Parramore (1997) supporting the use of cross-sectional data to test variables on one year. We have used a software data analysis package in excel to test the data along a multivariate analysis to obtain descriptive statistics of the total variables, such as mean, standard deviation, minimum, maximum, coefficient of variation, skewness and kurtosis.

---

[1] Data availability statement: https://www.bloomberg.com/professional/ allows to obtain the data; in our case it was obtained through the University of Leicester Library licence. Pertinent data can be obtained in a similar way by anyone concerned.

Next, we use the correlation method to estimate the relationship between independent, dependent, and control variables. Multiple linear regressions are "finally" employed to test corporate governance mechanisms on firms' financial performance.

### 3.2. Research and Sampling Design

The study uses the cross-sectional data method to test a sample of firms listed on London Stock Exchange for the year 2014 (LSE, 2014). The study is restricted to listed firms because they are expected to adhere to set regulation standards. In addition, listed firms are likely to prepare their accounting figures in compliance with international accounting practice (Ehikioya, 2009). We stress that we excluded financial institutions because they are subjects to different regulations from non-financial firms, whence may lead to outliers (Ausloos et al., 2018). In fact, mentioned scholars in prior reports have done likewise (Gust, 2009; Rodriguez-Fernandez, 2016). We have not taken into account the possibility of cross share holding (Rotundo and D'Arcangelis, 2010; D'Arcangelis and Rotundo, 2015; Cerqueti et al., 2018).

### 3.3. Hypotheses Development

After reviewing the literature from the prior studies, five hypotheses emerge:

*H1: Companies with large insider shareholding are those with superior financial performance* (Jensen and Meckling, 1976; McConnell and Servaes, 1990).

*H2: Companies with large board size achieve superior financial performance (*Anderson et al., 2004).

*H3: Companies displaying high proportion of board independence achieve high financial performance* (Beasley, 1996; Donaldson, 1990).

*H4: Companies with CEO duality achieve less financial performance* (Firstenberg and Malkiel, 1994)

*H5: Most high financial performance companies are those with high frequency audit committee meetings* (Kent and Stewart, 2008)

**3.4. Description of Variables and Measure Indicators**

This section covers descriptions of variables used in the study. These include dependent, independent and control variable. In addition, we will indicate measurement and proxies use to measure variables of corporate governance mechanism and their relationship with financial performance; see Table I. Selection of variables is based on prior literature (Christensen et al., 2010; Ehikioya, 2009) having considered both theoretical and empirical studies

3.4.1. Dependent Variables

Researchers have used various accounting –based measurement to estimate financial performance of companies (Christensen et al., 2010). These include sales, return on asset (ROE), earnings per share and growth. Accounting-based measures represent the historical figures focusing on management's stewardship of the company. However, these figures are sometimes distorted to suit management and might not represent the reality (Christensen et al., 2010). According to Core et al. (2006) operating profit measured by ROA is a better measure when examining the relationship between financial performance and corporate governance. For example, ROA is not affected by leverage, extraordinary items, and other discretionary items. In addition, other researchers (Brown and Caylor, 2009; Muth and Donaldson, 1998) have used ROA as a measure of accounting. Based on these factors and previous studies, we use ROA in this study.

Secondly, the forward-looking financial market measure Tobin's Q is used in this study. This is consistent with the efficient market hypothesis established by Malkiel and Fama (1970) where Tobin's Q was used to capture existing assets and future growth potentials of the company. Tobin's Q also captures investors' expectations to future events, including evaluation of current business strategies (Rose-Ackerman, 1973; Demsetz and Villalonga, 2001; Ehikioya, 2009; Christensen et al., 2010; Rodriguez-Fernandez, 2016). Let us describe the dependent variables:

**(i)    Return on Assets (ROA)**

The Return on Assets (ROA) gives an indication of how best the assets of a company is utilised to generate profit. The ROA is calculated by dividing annual earnings of the company by its total assets.

$$ROA = \frac{Net\ Income}{Total\ Assets} \times 100$$

**(ii)   Tobin's Q**

The Tobin's Q is a ratio of market value of company outstanding stock and debt divided by replacement cost of the company's assets ("book value") (Christensen et al., 2010).

$$Q\ Ratio = \frac{Total\ market\ value\ of\ firm}{Total\ Assets\ Value}$$

3.4.2. Independent Variables

The corporate governance mechanisms recalled in the empirical framework section are going to be the independent variables of this study. They are:

**(i)** **Insider Shareholding**

The Insider shareholding refers to any director, corporate officer or institutional investor who owns at least 10% of the total shares of a corporation (Jensen and Meckling, 1976). Insider ownership is measured as the percentage of company outstanding shares owned by such insiders:

$$Insider\ shareholding = \frac{Number\ of\ shares\ owned\ by\ insiders}{Total\ number\ of\ shares\ outstanding} \times 100$$

**(ii)** **Board Size**

The Company board size refers to the number of members on the board. There is some evidence to suggest that a large board size results in better decision making than a small board size thereby leading towards high financial performance (Williams et al., 2005).

$$Board\ size = Number\ of\ board\ of\ directors\ on\ company's\ board$$

**(iii)** **Independent Board**

The Independent board refers to outside board directors who are not affiliated to top executives of the firm (Fama and Jensen, 1983).

Independent board of directors can be estimated by dividing the number of non-executive directors by the total number of board of directors (multiplied by 100):

$$Independent\ board = \frac{Number\ of\ independent\ board}{Total\ number\ of\ board\ of\ directors} \times 100$$

**(iv) CEO Duality**

The CEO duality is when the CEO also holds the position of board chairperson. The role of the board of director is to monitor the CEO on behalf of shareholders. Corporate governance assumes a likelihood of concentration of power where the CEO plays dual roles (Christensen et al., 2010):

$$CEO\ duality = function\ of\ chair\ person\ combined\ with\ CEO$$

**(v) Audit Committee Meetings**

Audit committee meetings occur when the board of directors charged with the responsibility of financial reporting and disclosure of information for the company. It is argued by scholars that the frequency of audit committee meetings is strongly related to the performance of a company. The logic is that regular meetings will mean that more information can be obtained and disclosed (Christensen et al., 2010).

$$Audit\ committee\ meeting = Frequency\ of\ audit\ committee\ meetings$$

3.4.3. Control Variables

Researchers such as Christensen et al. (2010) and Ehikioya (2009) have used leverage and firm size as control variables in their study. The probable relevance has also been examined by Rodriguez-Fernandez (2016), Weir et al. (2002) and Essen et al. (2013). These variables are estimated through:

(i) $Firm\ size = Logarithm\ of\ total\ company\ assets$

(ii) $Leverage = \dfrac{Total\ Assets}{Total\ Shareholders\ Equity}$

## 2.5 Regression Models

Basing on prior studies by authors such as Guest (2009), Jackling and Johl (2009), Alfaraih et al. (2012), we propose two regression models to determine relations between good corporate governance mechanisms and financial performance of firms. The two model equations are

"**Model 1**":

$ROA = \beta_0 + \beta_1.IS + \beta_2.BS + \beta_3.IB + \beta_4.CD + \beta_5.AC + \beta_6.FS + \beta_7.LG + \varepsilon_i$

and similarly, "**Model 2**":

$Q\ Ratio = \gamma_0 + \gamma_1.IS + \gamma_2.BS + \gamma_3.IB + \gamma_4.CD + \gamma_5.AC + \gamma_6.FS + \gamma_7.LG + \eta_i$

**Table I: Summary of Variable Definition and its Measurement**

| Variable type | Variable Name | Definition and measurement |
|---|---|---|
| Dependent Variables | ROA | Return on assets, measured as net income/total assets × 100 |
| | Q ratio | Tobin's Q, measured as Total Market Value of Firm/Total Assets Value |
| Independent Variables | IS | Insider-shareholding, measured as the proportion of shares owned by insiders. |
| | BS | Board-size, measured as the number of board of directors on company's board. |
| | IB | Independent board, measured as proportion of independent board on company's board. |
| | CD | CEO duality, measured as a Function of board chair person combined with CEO, CEO = 1 if CEO is also chairperson, otherwise = 0. |
| | AC | Audit committee meetings, measured as a function of the number of audit committee meetings held. |
| Control Variables | FS | Firm size measured as the logarithm of the firm's total assets. |
| | LG | Leverage, measured as total assets/total shareholders' equity. |

## 4. Data Analysis and Discussion

This section contains the discussion of the empirical data use in the study. A correlation analysis is employed to show the relationship between corporate governance variables and either ROA or Tobin's Q. A regression analysis is presented to show how independent corporate governance mechanisms can either positively or negatively affect the dependent variables ROA or Tobin's Q.

### 4.1. Descriptive Statistics

The usual statistical characteristics, including mean, minimum, maximum, standard deviation, coefficient of variation (Cov.), skewness (Skew.), and kurtosis (Kurt.) are reported in Table II for the dependent, independent and control variables for a sample of 252 firms listed on London Stock Exchange data, extracted from Bloomberg.

ROA has a large standard deviation showing that the data is largely spread around the mean, whence the Coefficient of variation = 189%. This points to a high variation in the accounting-based performance among the UK's firms.

**Table II**: **Summary of Descriptive Statistics (N=252)**

| Variable | Min. | Max. | St. Dev. | Mean | Cov. | Skew. | Kurt. |
|---|---|---|---|---|---|---|---|
| ROA (%) | -0.68 | 0.54 | 0.11 | 0.056 | 189% | -1.92 | 18.07 |
| Q Ratio | 0.01 | 9.15 | 1.31 | 1.428 | 92% | 2.62 | 9.56 |
| IS (%) | 0.00 | 54.80 | 9.64 | 3.869 | 249% | 3.61 | 13.44 |
| BS (%) | 4.00 | 17.00 | 2.14 | 8.774 | 24% | 0.72 | 0.52 |
| IB | 0.21 | 0.92 | 0.12 | 0.625 | 20% | -0.49 | 0.30 |
| CD | 0.00 | 1.00 | 0.13 | 0.016 | 789% | 7.79 | 59.21 |
| AC | 0.00 | 14.00 | 1.78 | 4.504 | 40% | 1.76 | 5.87 |
| FS | 1.99 | 5.55 | 0.70 | 3.269 | 21% | 0.79 | 0.71 |
| LG | -33.59 | 46.96 | 4.89 | 2.948 | 166% | 2.82 | 44.92 |

For the Q ratio, the closeness of the mean and standard deviation signifies that the market-based performance among UK firms are closely netted. This is reflected in the relatively low coefficient of variation = 92%.

With reference to the independent variables, which represent corporate governance mechanisms, *IS* mean=3.9 and St. Dev.= 9.6 showing that the data is much distributed far from the mean: indeed, the Min and Max values are 0.0 and 54.8 respectively. This is reflected in a high coefficient of variation = 249%. This suggests that although some companies have about 2/3 of their shares held by insiders, 1/3 shares are still held by outsiders. The mean of *BS*=8.8 with a St. Dev. =2.1; Min=4, and Max=17 show that there are many similarities in firm's board size in UK.

Concerning the independent board of directors, the minimum and maximum values are 0.21 and 0.92 respectively; the mean= 0.63 and the standard deviation = 0.13 give a coefficient of variation = 20%. Thus, there are similar characteristics in the type of boards in UK firms. Most of the companies in the UK have independent board of directors perhaps because of transparency need and accountability associated with independent boards.

The CEO duality (*CD*) is characterized by a mean =0.02 and a standard deviation=0.13. This shows a huge deviation in the data spread of CEO duality, emphasized by a high coefficient of variation = 789%. These figures show that most firms in UK have separated the role of CEO and board chair person.

The statistical results for audit committee meetings held among UK firms show a low standard deviation=1.80 for the mean= 4.50, giving a low coefficient of variation = 40%. Notice that the (*AC*) data ranges from 0 (!) to a Max.= 14.

With reference to the control variables, the logarithm of assets has a mean = 3.3 and a standard deviation = 0.70 giving a low coefficient of variation = 21%. The minimum value = 1.10 and the maximum = 5.55. Here the data clusters around the mean, which implies a low size variation of such firms. The leverage shows a mean of 2.95 and standard deviation of 4.89 giving a coefficient of

variation = 166% in data spread. The data ranges from -33.59 to 46.96. Such a negative leverage implies that the cost of borrowing is greater than the return on investment. Thus, variations in firms' debt are rather consequent as confirmed by the 166% coefficient of variation value.

About the skewness, apart from ROA=-1.92 and *IB*=-0.49 that are negatively skewed, the remaining variable distributions present a positive skewness, indicating that the tail of these is on the right. Also in terms of kurtosis, with the exception of variables *CD*=59.21 and *LG*=44.92 that have heavy tail or outliers in the data distribution, the remaining variables have light tails or few outliers. These features point to an "interesting" random selection.

### 4.2. Correlation between variables

This section on correlations will help to determine whether there is multicollinearity among any of the variables. We noticed that prior researchers have raised concerns of possible multicollinearity among variables which could thereby distort the estimates of the regression results. In addition, because this research considers data for only one year period, there is no heteroscedasticity problem (Alin, 2010; Koop, 2008; Gujarity and Porter, 2009).

Table III displays the correlations between the dependent variable ROA and the independent variables and control variables, while their correlations with Tobin's Q are shown in Table IV.

**Table III. Correlation between ROA and Corporate Governance Mechanisms**

| Variable | ROA   | IS    | BS    | IB    | CD    | AC   | FS   | LG   |
|----------|-------|-------|-------|-------|-------|------|------|------|
| ROA      | 1.00  |       |       |       |       |      |      |      |
| IS       | 0.07  | 1.00  |       |       |       |      |      |      |
| BS       | 0.02  | -0.03 | 1.00  |       |       |      |      |      |
| IB       | 0.01  | -0.21 | 0.16  | 1.00  |       |      |      |      |
| CD       | 0.06  | 0.18  | 0.00  | -0.05 | 1.00  |      |      |      |
| AC       | -0.16 | -0.05 | 0.32  | 0.21  | 0.00  | 1.00 |      |      |
| FS       | -0.18 | -0.23 | 0.60  | 0.44  | 0.00  | 0.41 | 1.00 |      |
| LG       | 0.15  | -0.03 | 0.00  | 0.05  | -0.01 | 0.06 | 0.10 | 1.00 |

According to Gujarity and Porter (2009), a correlation above 0.8 signals a possible evidence of multicollinearity in the data set. The results in Table III indicate that multicollinearity is unlikely to be a problem here. Nevertheless, there are mixed results: several variables are negatively correlated to ROA, whereas others have positive correlation with *ROA*. "Interestingly", the correlation is negative for *FS*=-0.18 and *AC*=-0.16. All other variables have positive correlation with *ROA*: *IS*=0.07, *BS*=0.02, *IB*=0.01, *CD*=0.06 and *LG*=0.15. This suggests that an increase in any of these variables increase with *ROA*, whereas variables *FS* and *AC* decrease with *ROA*.

Concerning Tobin's Q and corporate governance mechanisms (Table IV), again, multicollinearity is unlikely to be a problem as none of the variables has a correlation above 0.8. Furthermore, only variables, *IS* and *CD* have a positive correlation with Tobin's Q (0.20, 0.01) respectively, suggesting that when these corporate governance variables increase, Tobin's Q increases also. The variables *BS* (0.05), *IB* (-0.19), *AC* (-0.16), *FS* (-0.42), and *LG* (-0.09) have a negative correlation with Tobin's Q, suggesting that these variables decrease with Tobin's Q.

**Table IV. Correlation between Q ratio and Corporate Governance Mechanisms**

| Variable | Q ratio | IS | BS | IB | CD | AC | FS | LG |
|---|---|---|---|---|---|---|---|---|
| Q ratio | 1.00 | | | | | | | |
| IS | 0.20 | 1.00 | | | | | | |
| BS | -0.05 | -0.03 | 1.00 | | | | | |
| IB | -0.19 | -0.21 | 0.16 | 1.00 | | | | |
| CD | 0.01 | 0.18 | 0.00 | -0.05 | 1.00 | | | |
| AC | -0.16 | -0.05 | 0.32 | 0.21 | 0.00 | 1.00 | | |
| FS | -0.42 | -0.23 | 0.60 | 0.44 | 0.00 | 0.41 | 1.00 | |
| LG | -0.09 | -0.03 | 0.00 | 0.05 | -0.01 | 0.06 | 0.10 | 1.00 |

## 4.3. Regression Analysis

In this section, the multiple linear regressions models, see Sect. 2.5, are used in order to establish the impact of corporate governance mechanisms on the two response variables, ROA and Tobin's Q. The statistical results are given in Table V and Table VI respectively.

Table V points to mixed results between variables and their impacts on financial performance. A few predictor variables are statistically significant, while others are not. For instance, p values for *IS* (0.81) and *CD* (0.31) are high, considering a significant level of 0.00. This implies that the above-mentioned variables are not statistically significant and do not have predictive power on *ROA*. Therefore, changes in *IS* and *CD* will not have any impact on the financial performance of firms when measured through *ROA*.

However, variables *BS* (0.00), *IB* (0.03), *AC* (0.04), *FS* (0.00), and *LG* (0.00) have low p-values, which imply a predictive power on the *ROA*. As such, the regression Model 1 could be reduced to

$$ROA = 0.012\, BS + 0.127\, IB - 0.008\, AC - 0.008\, FS + 0.004\, LG + \varepsilon_i$$

It should be noted that the intercept value $\beta_0 = 0$, because of an insignificant p-value, means that the intercept is not significantly different from 0. Practically, from the reduced Model 1 equation, a prediction can be made that, for any additional change in *BS*, one can expect ROA to increase on average by 12%. However, additional change in *AC*, will result that *ROA* on average would decrease by -0.8 %, because of the negative coefficient.

In addition, $R^2 = 0.12$, see Table V, implies that only 12% of all the independent and control variables explain the effects on the dependent variable *ROA*, whence 88% of ROA behaviour has to be explained by other independent variables not included in this study.

**Table V: Multivariate Regression Results for ROA**

| Multiple Regression Results for ROA | | | |
|---|---|---|---|
| Variable | Coefficients (x 100) | t-test | p-value |
| Intercept | 6.6332 | 1.6208 | 0.11 |
| IS | 0.0172 | 0.2451 | 0.81 |
| BS | 1.2385 | 3.2405 | 0.00 |
| IB | 12.678 | 2.1858 | 0.03 |
| CD | 5.3167 | 1.0269 | 0.31 |
| AC | -0.8376 | -2.1209 | 0.04 |
| FS | -5.3093 | -3.9690 | 0.00 |
| LG | 0.3918 | 2.9896 | 0.00 |
| *Regression Statistics* | | | |
| $R^2$ | | | 0.120 |
| Observations | | | 252 |

*Significant level: 0.000*

From Table VI, p-values for *IS* (0.15), *IB* (0.42), *CD*, *AC* (0.68), and *LG* (0.53) are found to be high, which implies a lack of predictive power on Tobin's Q; these variables could be removed from Model 2. In contrast, two independent variables *BS* (0.00) and *FS* (0.00) are statistically significant and do have predictive power on Tobin's Q since they have p-values close to 0. The new equation for a reduced Model 2 can be rewritten as follows:

$$Q\ Ratio = 3.18 + 0.19\ BS - 1.13\ FS + \eta_i$$

**Table VI: Multivariate Regression Results Summary for Q Ratio**

| Multiple Regression Results of Q ratio | | | |
|---|---|---|---|
| *Variable* | *Coefficients* | *t –test* | *p-value* |
| Intercept | 3.1817 | 6.8030 | 7.86 E-11 |
| *IS* | 0.0116 | 1.4471 | 0.1492 |
| *BS* | 0.1901 | 4.3535 | 1.97 E-05 |
| *IB* | 0.5380 | 0.8117 | 0.4178 |
| *CD* | -0.0555 | -0.0938 | 0.9254 |
| *AC* | -0.0186 | -0.4130 | 0.6800 |
| *FS* | -1.1288 | -7.3839 | 2.42 E-12 |
| *LG* | -0.0094 | -0.6298 | 0.5294 |
| *Regression Statistics* | | | |
| $R^2$ | | | 0.253 |
| Observations | | | 252 |

*Significant level: 0.000*

Therefore, one can make a prediction that for a unit increase in *BS*, holding all other factors constant, the Q ratio on average will increase by 19%. However, all things being equal, a unit increase in *FS* will have a corresponding average decrease equal to -113 % of the Q ratio.

An important feature should be emphasized: the independent variable *IS* is not significant in either models. This means that insider shareholding does not influence financial performance (measured through *ROA* and Tobin's *Q*). This confirms the findings of Agrawal and Knoboer (1996). However, this finding disagrees with Jensen and Meckling (1976). Therefore, our first hypothesis that companies with a large insider shareholding are those with superior financial performance can be rejected.

*A contrario*, *BS* is statistically significant in both models, implying that a large board size could improve financial performance (*ROA* and Tobin's Q). This finding is consistent with Anderson et al. (2004). Therefore, our second hypothesis is acceptable.

Concerning *IB*, which presents a statistically significant effect in *ROA*, one concludes that additions to the board of independent directors will improve financial performance. This is consistent with Beasley (1996) and Donaldson (1990). One could consider that the third hypothesis can be accepted concerning ROA. However, in terms of Tobin's Q, *IB* is statistically insignificant: the independent board of directors has no effect on financial performance. One should reject the third hypothesis for Tobin's Q.

In terms of *CD*, there is a lack of statistical significance for both dependent variables (*ROA* and Tobin's Q), suggesting that it does not matter whether there is a dual role or a separation of CEO and chair role: the financial performance remains unaffected. This finding is inconsistent with Firstenberg and Malkiel (1994) who suggest that companies with CEO duality do not perform well financially. Therefore, the fourth hypothesis is rejected.

Regarding *AC*, a statistical significance is obtained for *ROA* but with a negative coefficient. This suggests that increasing the frequency of audit committee meetings impacts negatively on the financial performance *(ROA)*. This finding disagrees with that of Kent and Stewart (2008). However, in terms of Tobin's Q no statistical significance is found, which suggests that the frequency of audit committee meetings lacks some predictive ability on the financial performance (Tobin's Q). This finding is consistent with Weir et al. (2002). Therefore, the fifth hypothesis is also rejected.

## 5. Discussion

The above results explain why there is a controversy in this field of study. The statistics so obtained provide mixed results, depending on the Model. There are corporate governance mechanisms that have no statistical significance; some have positive, and others have a negative statistical significance on estimating financial performance, we stress, using *ROA* or Tobin's *Q*.

The findings show that, insider shareholding has insignificant influence on both ROA and Tobin's Q. This supports the findings of Agrawal and Knoeber (1996). The implication is that whether managers own many or a few shares in a company is irrelevant for the financial performance. This means, there should be no hindrance to pay executive bonuses in shares instead of salary, in order to increase insider shareholding.

One field of controversy is board size and its impact on financial performance. The outcome of the study indicates a positive statistic significance of board size on the two financial performance ratios (ROA and Tobin's Q). It is seen that increasing the size of the board improves financial performance contrary to the argument of Jensen (1993). The findings in this study, however, support the argument of Anderson et al. (2004) that large boards help in proper allocation of committee work for enhancing growth and financial performance. This argument supports the views of Fama and Jensen (1983) who argued that the role of the board involves monitoring managerial behaviour, which is likely to be more effective with a large board size. In this respect, one can follow Williams et al.

(2005) arguing that financial markets place a high premium on large board size, perceived to be better resourced for monitoring and or skills transfer abilities.

Both Donaldson (1990) and Beasley (1996) argued in favour of a high proportion of board independent members. They documented that companies displaying high proportion of board independent directors achieve a high financial performance. For instance, Donaldson (1990) stressed that those companies command creditability in accounting, whence investors seeing this have a favourable opinion. The outcome of our study suggests that there is a statistical significance of an independent board on ROA, as in Donaldson (1990) and Beasley (1996). However, an insignificant test result is discovered for board independent influence on the Q ratio, as in Fosberg (1989). Thus, in terms of the Q ratio, one can suggest that companies should not be concerned by board characteristics, either executive or non-executive.

We have pointed out the evidence from the regression results about the lack of statistical significance of CEO duality on both financial performance indicators (ROA and Tobin's Q). This finding is inconsistent with Fistenberg and Malkiel (1994) who documented that CEO duality has a negative impact. From this finding, we consider that firms might save some money by employing one person as CEO and chairperson instead of two persons, - but that should be locally discussed.

With reference to audit committee meetings, different results were obtained for ROA and Tobin's *Q*. Significant results are obtained for ROA supporting the views of Kent and Stewart (2008) that a high frequency of audit committee meetings encourages high financial performance. However, insignificant statistical results are obtained for Tobin's Q, indicating that the number of audit committee meetings does not matter: financial performance will remain unchanged. This supports the findings of Weir et al. (2002).

Finally, considering R-square, only 12% and 25% of the response variables explained ROA and Tobin's Q variation. Thus, several variables appear not to be included for explaining ROA and Tobin's Q. Therefore, some further imagination and studies are needed by researchers about this theoretical deficit.

## 5. Conclusion

### 5.1. Summary and Concluding Remarks

This study has examined the impact of 5 corporate governance mechanisms (insider shareholding, board size, independent directors, CEO duality, and audit committee meetings) on financial performance (ROA and Tobin's Q), taking into account 2 control variables. The study covers a sample of 252 firms listed on London Stock Exchange in 2014. Two theories of corporate governance, agency theory, and stewardship theory, form the theoretical framework. The outcome of the regression results displays mixed findings similar to prior studies (Christensen et al., 2010; Rodriguez-Fernandez, 2016).

For instance, many prior studies suggest that the size of insider shareholding affects the financial performance (Jenson and Meckling, 1976: Fama and Jensen, 1983; Gupta and Sachdeva, 2017). However, the outcome indicates that insider shareholding has no influence on financial performance, itself consistent with findings of Agrawal and Knoeboer (1996).

Some of the corporate governance mechanisms such as board size and independent board members exhibited predictive power on both financial performance indicators, ROA and Tobin's Q. This finding is in agreement with Christensen et al. (2010) further concluding that a strong independent board is one of the solutions to agency problem by reducing cost, thereby improving financial performance. Somewhat inconclusively, the frequency of audit committee meetings indicates some influence on the financial performance indicator ROA but no influence on Tobin's Q.

Finally, our study about CEO duality demonstrated no influence on both ROA and Tobin's Q. This finding disagrees with the conflicting prior literatures having examined the variable. The supporters of agency theory suggest a positive outcome when the role of the CEO and the chairperson is separated (Rechner and Dalton, 1991; Balatbat *et al,* 2004). On the other hand, supporters of stewardship theory argue for role of the CEO and the chairperson to be combined as it allows clear leadership direction that improves performance (Stoeberl and Sherony, 1985). However, our finding demonstrates a neutral cause; it does not matter whether the CEO and chairperson's role is combined, or otherwise; the outcome of financial performance remains unchanged whatever the board choice.

To help achieve a robust finding, the corporate governance mechanisms were controlled by firm size and leverage. For the firm size, the regression coefficients are negative, but have opposite values as concerns the leverage. Notice that we have not taken into account the possibility of cross share holding even though some thought should be given on the matter (Rotundo and D'Arcangelis, 2010; D'Arcangelis and Rotundo, 2015; Cerqueti et al., 2018), investigating "cross performances".

Thus, our study, like previous studies, provides mixed findings and partial conclusions to the debate. However, it has strengthened some of the existing theoretical framework. We can conclude that companies in the United Kingdom can improve their financial market performance by adopting the right corporate governance mechanisms. We have indicated which (among others) corporate governance mechanisms influence financial performance indeed.

Thus, future researchers could explore other theories like stakeholder theory, shareholder theory, leadership cycle theory and others in order to introduce other variables in the considerations, such as board diligence or CEO tenure. In addition, other factors such as technology, global financial crises, economic conditions (booms and recessions), cross share holdings, … can be investigated, as they might likely have some impact on financial performance. Furthermore, panel data can be employed to test variables over several years with other data sizes. Brexit influence will likely attract new considerations.

## 6. Acknowledgements

**Data Availability statement:**

The data that support the findings of this study are available from [Bloomberg : https://www.bloomberg.com/professional/ ]
Restrictions apply to the availability of these data, which were used under license, at the University of Leicester Library, for this study. Data are available from the authors with the permission of University of Leicester and Bloomberg

# 7. References

Agrawal, A. and Knoeber, C. R. (1996) Firm Performance and Mechanisms to Control Agency Problems between Managers and Shareholders. *Journal of Financial and Quantitative Analysis,* 31(3), pp. 377-397.

Aguilera, R. V. (2005) Corporate Governance and Director Accountability: An institutional comparative perspective. *British Journal of Management,* 16(1), pp. 539-553.

Akshita, A. and Sharma, C.(2015) Impact of Firm Performance on Board Characteristics: Empirical Evidence from India. *IIM Kozhikode Society & Management Review,* 4(1), pp. 53–70.

Alfaraih, M., Alanezi, F. and Almujamed, H. (2012) The Influence of Institutional and Governance Ownership on Financial Performance: Evidence from Kuwait. *Institutional Business Research*, 5(10), pp. 192-200.

Alin, A. (2010) Multicollinearity. *Wiley Interdisciplinary Reviews: Wires Computational Statistics,* 2(3), pp. 370-374.

Allen, F. and Gale, D.(2001) *Comparing Financial Systems.* Cambridge, MA: MIT Press.

Al-Najjar, B. (2017) Corporate Governance and CEO Pay: Evidence from UK Travel and Leisure listed firms. *Tourism Management,* 60, pp. 9-14.

Anderson, R. C., Mansib, S. A. and Reeb, D. (2004) Board characteristics, accounting report integrity, and the cost of debt. *Journal of Accounting and Economics,* 37(3), pp. 315-342.

ASX Corporate Governance Council (2007) *Corporate Governance Principles and Recommendations.* 2nd edition ed. Sydney: ASX.

Ausloos, M., Bartolacci, F., Castellano, N. G., & Cerqueti, R. (2018). Exploring how innovation strategies at time of crisis influence performance: a cluster analysis perspective. *Technology Analysis & Strategic Management*, *30*(4), pp. 484-497.

Balatbat, M.C.A., Taylor, S. L. and Walter, T.S. (2004) Corporate Governance, Insider Ownership and Operating Performance of Australian Initial Public Offerings. *Accounting and Finance,* 44(3), pp. 299-328.

Beasley, M. S. (1996) An Empirical Analysis of the Relation between the Board of Director Composition and Financial Statement Fraud. *The Accounting Review,* 71(4), pp. 443-465.

Bonazzi, L. and Islam, S. M. N. (2007) Agency Theory and Corporate Governance: A Study of the Effectiveness of Board in their Monitoring of the CEO. *Journal of Modelling in Management,* 2(1), pp. 7-23.

Boyd, B. K. (1995) CEO Duality and Firm Performance: A contingency model. *Strategic Management Journal,* 16(4), pp. 301-312.

Brown, L.D and Caylor, L.C. (2009) Corporate Governance and Financial Operating Performance. Review of Quantitative Finance and Accounting, 32(2), pp. 129-144.

Cadbury, A. (2000) The Corporate Governance Agenda. *Corporate Governance: An International Review,* 8(1), pp. 7-15.

Cannella, A.A. and Lubatkin, M. (1993) Succession as a Sociopolitical Process: Internal Impediments to Outsider Selection. *Academy of Management Journal,* 36(4), pp. 763-793.

Cerqueti, R., Rotundo, G., and Ausloos, M. (2018). Investigating the configurations in cross-shareholding: a joint copula-entropy approach. *Entropy,* 20(2), 134.

Christensen, J., Kent, P. and Stewart, J. (2010) Corporate Governance and Company Performance in Australia. *Australian Accounting Review,* 20(4), pp. 372-386.

Clarke, T. (2004) *Theories of corporate governance : the philosophical foundations of corporate governance.* London: Routledge.

Clarke, T. (2007) *International corporate governance. A comparative approach.* London: Routledge.

Core, J. E., Guay, W. and Rusticus, T. (2006) Does Weak Governance Cause Weak Stock Returns? An Examination of Firm Operating Performance and Investors' Expectations. *The Journal of Finance,* 61(2), pp. 655-687.

D'Arcangelis, A. M., and Rotundo, G. (2015). Mutual funds relationships and performance analysis. *Quality & Quantity*, 49(4), pp. 1573-1584.

Davidson, R., Goodwin-Stewart, J. and Kent, P. (2005) Internal Governance Structures and Earnings Management. *Accounting and Finance,* 45(2), pp. 241-267.

Deakin, S. and Konzelmann, S. J. (2004) Learning from Enron. *Corporate Governance: An International Review,* 12(2), pp. 134-142.

Demsetz, H. and Villalonga, B. (2001) Ownership Structure and Corporate Performance. *Journal of Corporate Finance,* 7(3), pp. 209-233.

Donaldson, L. (1990) The Ethereal Hand: Organisational Economics Theory. *Academy of Management Review,* 15(3), pp. 369-381.

Donaldson, L. and Davis, J.H. (1991) Stewardship Theory or Agency Theory: CEO Governance and Shareholder Returns. *Australian Journal of Management,* 16(1), pp. 49–64.

Ehikioya, B. I. (2009) Corporate Governance Structure and Firm Performance in Developing Economies: Evidence from Nigeria. *The international Journal of Business in Society,* 9(3), pp. 231-243.

Elsayed, K. (2007) Does CEO Duality Really Affect Corporate Performance?. *Corporate Governance: an International Review,* 15(6), pp. 1203-1214.

Essen, M., Engelen, P. and Carney, M. (2013) Does "Good" Corporate Governance Help in a Crisis?. *Corporate Governance: An International Review,* 21(3), pp. 201-224.

Fama, E. (1980) Agency Problems and the Theory of the Firm. *Journal of Political Ecoconomics,* 88(2), pp. 288-307.

Fama, E. F. and Jensen, M. (1983) Separation of Ownership and Control. *Journal of Law and Economics,* 26(2), pp. 301-3025.

Firstenberg, P. B. and Malkiel, B. G. (1994) The Twenty-First Century Boardroom: Who will be in charge? *Sloan Management Review,* 36(1), pp. 27-35.

Florackis, C. (2005) Internal Corporate Governance Mechanisms and Corporate Performance: Evidence for UK firms. *Applied Financial Economics Letters,* 1(4), pp. 211-216.

Fosberg, R. H. (1989) Outside Directors and Managerial Monitoring. *Akron Business and Economic Review,* 20(2), pp. 24-32.

Freeman, R.E., (1983). Strategic Management: A stakeholder approach. *Advances in Strategic Management*, *1*(1), pp. 31-60.

Freeman, R.E., (2010). *Strategic Management: A stakeholder approach*. Cambridge University Press.

Gill, A. (2008) Corporate Governance as Social Responsibility: A Research Agenda. *Berkeley Journal of International Law,* 26(2), pp. 452-478.

Gillan, S. L. (2006) Recent Development in Corporate Governance: An Overview. *Journal of Corporate Finance,* 12(3), pp. 381-402.

Guest, P. M. (2009) The Impact of Board Size on Firm Performance: Evidence from the UK. *The European Journal of Finance,* 15(4), pp. 385-404.

Gujarati, D. N. and Porter, D. (2009) *Basic Econometrics.* Boston: McGraw-Hill.

Gupta, A. and Sachdeva, K. (2017) Skin or Skim? Inside Investment and Hedge Fund Performance. *NYU Working Paper No. 2451/38717,* p. 62.

Haniffa, R. and Hudaib, M. (2006) Corporate Governance Structure and Performance of Malaysian Listed Companies. *Journal of Business Finance and Accounting,* 33(7-8), pp. 1034–1062.

Jackling, B. and Johl, S. (2009) Board Structure and Firm Performance: Evidence from India's Top Companies. *Corporate Governance: An International Review,* 17(4), pp. 492–509.

Jensen, C. M. and Meckling, H. W. (1976) Theory of firm:Managerial Behaviour, Agency Costs and Ownership Structure. *Journal of Financial Economics,* 3(4), pp. 305-360.

Jensen, M. C. (1993) The Modern Industrial Revolution, Exit, and the Failure of Internal Control Systems. *The Journal of Finance,* 48(3), pp. 831–880.

Kent, P. and Stewart, J. (2008) Corporate Governance and Disclosures on the Transition to International Financial Reporting Standards. *Accounting and Finance,* 48(4), pp. 649-71.

Klein, A. (1998) Firm Performance and Board Committee Structure. *The Journal of Law and Economics,* 41(1), pp. 275-304.

Koop, G. (2008) *Introduction to Econometrics.* West Sussex: John Wiley and Sons Ltd.

Lipton, M. and Lorsch, J. W. (1992) A Modest Proposal for Improved Corporate Governance. *The Business Lawyer,* 48(1), pp. 59-77.

Lorsch, J. W. and MacIver, E. (1989) *Pawns or Potentates: The Reality of America's Corporate Boards.* Boston MA: Harvard Business School Press.

LSE (2014) see: https://en.m.wikipedia.org/wiki/FTSE_100_Index;

https://www.londonstockexchange.com/exchange/prices-and-markets/stocks/indices/summary/summary-indices-constituents.html?index=UKX, for the selection of firms.

Moreover, the London Stock Exchange (LSE) has a Historic Price Service (HPS) on the website

http://www.londonstockexchange.com/products-and-services/reference-data/hps/hps.htm

providing quotations for all securities traded on the London Stock Exchange since 1999. The data was obtained through the University of Leicester licence and

https://www.bloomberg.com/professional/

Malkiel, B. G., & Fama, E. F. (1970). Efficient capital markets: A review of theory and empirical work. *The journal of Finance*, *25*(2), pp. 383-417.

Mallin, C. A. (2016) *Corporate Governance.* 5th ed. Oxford: Oxford University Press.

McConnell, J. J. and Servaes, H. (1990) Additional Evidence on Equity Ownership and Corporate Value. *Journal of Financial Economics,* 27(2), pp. 595-612.

Mura, R. (2007) Firm Performance: Do Non-Executive Directors Have Minds of their Own? Evidence from UK Panel Data. *Financial Management,* 36(3), pp. 81–112.

Muth, M. and Donaldson, L. (1998) Stewardship Theory and Board Structure: A Contingency Approach. *Corporate Governance: An International Review,* 6(1), pp. 5-28.

Nicholson, G. J. and Kiel, G. C. (2007) Can Directors Impact Performance? A Case-Based Test of Three Theories of Corporate Governance. *Corporate Governance: An International Review,* 15(4), pp. 585–608.

Pamela, K. and Jenny, S. (2008) Corporate Governance and Disclosures on the Transition to International Financial Reporting Standards. *Accounting & Finance,* 48(4), pp. 649-671.

Pankaj, K. and Zabihollah, R. (2006) The Sarbanes-Oxley Act of 2002 and Capital-Market Behavior: Early Evidence. *Contemporary Accounting Research,* 23(3), pp. 629–654.

Perfect, S. B. and Wiles K.W. (1994) Alternative Construction of Tobin's Q. An empirical comparison. *Journal Empire of Finance,* 1(3-4), pp. 313-341.

Rechner, P.L. and Dalton, D.R. (1991) CEO Duality and Organisational Performance: A Longitudinal Analysis. *Strategic Management Journal,* 12(2), pp. 155-160.

Rodriguez-Fernandez, M. (2016) Social responsibility and financial performance: The role of good corporate governance. *BRQ Business Research Quarterly*, *19*(2), pp.137-151.

Rose-Ackerman, S. (1973) Effluent Charges: A Critique. *The Canadian Journal of Economics,* 6(4), pp. 512-528.

Rosenstein, S. and Wyatt, J. G. (1990) Outside Directors, Board Independence, and Shareholder Wealth. *Journal of Financial Economics,* 26(2), pp. 175-191.

Rotundo, G., and D'Arcangelis, A. M. (2010). Network analysis of ownership and control structure in the Italian Stock market. *Advances and Applications in Statistical Sciences,* Special Issue 2(2), pp. 255-274.

Stanwick, P. A. and Stanwick, S. D. (2002) The Relationship Between Corporate Governance and Financial Performance: An Empirical Study. *Journal of Corporate Citizenship,* 8(4), pp. 35-48.

Stoeberl, P. A and Sherony, B. C. (1985) *Board Efficiency and Effectiveness. Handbook for Corporate Directors.* New York: McGraw-Hill.

Terjesen, S., Couto, E.B. and Francisco, P.M., (2016). Does the presence of independent and female directors impact firm performance? A multi-country study of board diversity. *Journal of Management & Governance*, *20*(3), pp. 447-483.

Turnbull, S.(1997) Corporate Governance: Its Scope, Concerns and Theories. *Corporate Governance: An International Review,* 5(4), pp. 180-205.

Vafeas, N. and Theodorou, E. (1998) The Relationship between Board Structure and Firm Performance in the UK. *The British Accounting Review,* 30(4), pp. 383-407.

Watsham, T. J. and Parramore, K. (1997) *Quantitative Methods in Finance*. London: Thomson.

Weir, C. (1997) Acquisitions and Firm Characteristics: The Importance of Internal Monitoring Mechanisms. *Management Decision,* 35(2), pp. 155-162.

Weir, C., Laing, D. and McKnight, P. J. (2002) Internal and External Governance Mechanisms: Their Impact on the Performance of Large UK Public Companies. *Journal of Business Finance and Accounting,* 29(5-6), pp. 579–611.

Williams, R., Fadil, P. and Armstong, R. (2005) Top Management Team Tenure and Corporate Illegal Activity: The Moderating Influence of Board Size. *Journal of Managerial Issues,* 17(4), pp. 479-493.

Yekini, K.C., Adelopo, I., Andrikopoulos, P. and Yekini, S., (2015) Impact of board independence on the quality of community disclosures in annual reports. *Accounting Forum* 39(4), pp. 249-267.

Yermack, D. (1996) Higher market valuation of companies with a small board of directors. *Journal of Financial Economics,* 40(2), pp. 185-211.